\documentclass[fleqn,usenatbib]{mnras}

\usepackage{newtxtext,newtxmath}
\usepackage[T1]{fontenc}

\DeclareRobustCommand{\VAN}[3]{#2}
\let\VANthebibliography\thebibliography
\def\thebibliography{\DeclareRobustCommand{\VAN}[3]{##3}\VANthebibliography}

\usepackage{graphicx}	% Including figure files
\usepackage{amsmath}	% Advanced maths commands

\usepackage{graphicx}
 \usepackage{setspace}
\usepackage{textcomp}     % access \textquotesingle
\usepackage{url}
\usepackage[]{amsmath}
\usepackage{amsmath}
\usepackage{nccmath}
\usepackage{lineno}
\usepackage[flushleft]{threeparttable}
\newcommand{\an}{{\textquotesingle Ayl\'{o}\textquotesingle chaxnim }}

\newcommand{\psj}{{Planetary Science Journal}}

\title[Discovery and characterization of C/2022 E3]{Palomar discovery and initial characterization of naked-eye long period comet C/2022 E3 (ZTF)}

\author[B. T. Bolin et al.]{
B. T. Bolin,$^{1,2,3}$\thanks{NASA Postdoctoral Program Fellow}$^{,}$\thanks{E-mail: bryce.bolin@nasa.gov (BTB)}
F. J. Masci,$^{3}$
D. A. Duev,$^{4}$
J. W. Milburn,$^{5}$
J. D. Neill,$^{2,5}$
\newauthor
 \;J. N. Purdum,$^{5}$
C. Avdellidou,$^{6}$
Y.-C. Cheng,$^{8,9}$
M. Delbo,$^{6}$
C. Fremling,$^{2,5}$
\newauthor
 \;M. Ghosal,$^{10}$
Z.-Y. Lin$^{11}$,
C. M. Lisse$^{12}$,
A. Mahabal$^{2}$,
and M. Saki$^{7,13}$
\\
$^{1}$Goddard Space Flight Center, 8800 Greenbelt Road, Greenbelt, MD 20771, USA\\
$^{2}$Division of Physics, Mathematics and Astronomy, California Institute of Technology, Pasadena, CA 91125, USA\\
$^{3}$Infrared Processing and Analysis Center, California Institute of Technology, Pasadena, CA 91125, USA\\
$^{4}$Weights $\&$ Biases, San Francisco, CA 94103, USA\\
$^{5}$Caltech Optical Observatories, California Institute of Technology, Pasadena, CA 91125, USA\\
$^{6}$Universit\'{e} C\^{o}te d'Azur, CNRS-Lagrange, Observatoire de la C\^{o}te d'Azur, Nice 06304, France\\
$^{7}$Department of Physics, Edmund C. Leach Science Center, Auburn University, Auburn, AL 36849, USA\\
$^{8}$Department of Physics, National Taiwan Normal University, Taipei City 116325, Taiwan\\
$^{9}$Center of Astronomy and Gravitation, National Taiwan Normal University, Taipei City 116325, Taiwan\\
$^{10}$Institute for Astronomy, University of Hawai'i, 2680 Woodlawn Dr., Honolulu, HI 96822, USA\\
$^{11}$Institute of Astronomy, National Central University, Taoyuan City 32001, Taiwan\\
$^{12}$Johns Hopkins University Applied Physics Laboratory, 11100 Johns Hopkins Rd, Laurel, MD 20723, USA\\
$^{13}$Department of Mathematics, Physics, Astronomy, $\&$ Statistics, University of Missouri, St Louis, MO 63121, USA\\
}
\date{Accepted XXX. Received YYY; in original form ZZZ}

\pubyear{2023}

\begin{document}
\label{firstpage}
\pagerange{\pageref{firstpage}--\pageref{lastpage}}
\maketitle

% Abstract of the paper
\begin{abstract}
Long-period comets are planetesimal remnants constraining the environment and volatiles of the protoplanetary disc. We report the discovery of hyperbolic long-period comet C/2022 E3 (ZTF), which has a perihelion $\sim$1.11 au, an eccentricity $\gtrsim$1 and an inclination $\sim$109$^{\circ}$, from images taken with the Palomar 48-inch telescope during morning twilight on 2022 Mar 2. Additionally, we report the characterization of C/2022 E3 (ZTF) from observations taken with the Palomar 200-inch, the Palomar 60-inch, and the NASA Infrared Telescope Facility in early 2023 February to 2023 March when the comet passed within $\sim$0.28 au of the Earth and reached a visible magnitude of $\sim$5. We measure g-r = 0.70$\pm$0.01, r-i = 0.20$\pm$0.01, i-z = 0.06$\pm$0.01, z-J = 0.90$\pm$0.01, J-H = 0.38$\pm$0.01 and H-K = 0.15$\pm$0.01 colours for the comet from observations. We measure the A(0$^\circ$)f$\rho$ (0.8~$\mu$m) in a 6500~km radius from the nucleus of 1483$\pm$40~cm, and CN, C$_3$, and C$_2$ production of 5.43$\pm0.11\times$10$^{25}$~mol/s, 2.01$\pm0.04\times$10$^{24}$, and 3.08$\pm0.5\times$10$^{25}$~mol/s, similar to other long period comets. We additionally observe the appearance of jet-like structures at a scale of $\sim$4,000 km in wide-field g-band images, which may be caused by the presence of CN gas in the near-nucleus coma.
\end{abstract}

% Select between one and six entries from the list of approved keywords.
% Don't make up new ones.

%%%%%%%%%%%%%%%%%%%%%%%%%%%%%%%%%%%%%%%%%%%%%%%%%%

%%%%%%%%%%%%%%%%% BODY OF PAPER %%%%%%%%%%%%%%%%%%

\section{Introduction}
Long-period comets, defined as having an orbital period $>$200 years, originate from the Oort Cloud, an isotropic reservoir of nuclei extending to 10,000's of au from the Sun \citep[][]{Oort1950}. It is modeled that the Oort cloud was created by the scattering of planetesimals in the original trans-Neptunian disc onto orbits with semi-major axes $>$10,000 au through close encounters with the gas giants \citep[][]{Duncan1987,Vokrouhlicky2019}. Interactions with galactic tides and encounters with passing stars can cause the perihelia of the scattered comets to cross the planetary region where they are scattered into the inner solar system by planetary encounters \citep[e.g.,][]{Dybczynski2022}.

While planetesimals in the original trans-Neptunian disc are thought to be initially formed as large 100 km-scale bodies \citep[][]{Klahr2020,Simon2022}, collisional evolution subsequent to their formation resulted in the production of collisional aggregates $\lesssim$10 km in size \citep[][]{Bottke2023}. Therefore, the Oort Cloud was populated by many 10 km-scale or smaller bodies that have returned as long-period comets as detected by ground-based surveys \citep[][]{Boe2019} providing the opportunity to study the remnants of the original planetesimals.

The Palomar Observatory 48-inch (P48) telescope's Zwicky Transient Facility (ZTF) camera scans the observable night sky twice every night \citep[][]{Bellm2019} and has the capability of observing asteroid and comet transients down to V$\sim$20-21 \citep[e.g.,][]{Bolin2021LD2, Farnocchia2022,Bolin2022IVO}. The ZTF survey time is divided between a general all-sky survey \citep[][]{Graham2019} and micro-surveys specifically designed to detect solar system objects \citep[][]{Chang2022,Bolin2023Com}. Both survey types are processed with data analysis techniques designed to discovery and characterize asteroids and comets \citep[][]{Masci2019,Duev2019,Duev2021,Golovich2021}. On 2022 March 2, a cometary object, C/2022 E3 (ZTF), was discovered in P48/ZTF images taken during morning twilight \citep[][]{Bolin2022E3}.

Operation of telescopes such as the Palomar 60-inch telescope (P60), Palomar 200-inch (P200), and the Infrared Telescope Facility have been used to follow up and characterize asteroids and comets observed by the P48-inch \citep[][]{Bolin20202I,Purdum2021,Bolin2023NT}. Following the initial observations of the comet (hereafter E3) P48/ZTF, the comet was observed by the Palomar 200-inch telescope, the Infrared Telescope Facility (IRTF) and the Palomar 60-inch telescope (P60) in order to characterize its physical properties. We will describe the P48 and the follow-up observations with the P200, P60, and IRTF in greater detail in the following section.

\section{Observations}
In addition to its all-sky coverage, the P48/ZTF scans the portions of the night sky within 60 of the Sun during nautical and astronomical twilight each morning and evening to search for Earth co-orbitals \citep[][]{Yeager2022,Yeager2023}, comets \citep[][]{Duev2021}, Atira asteroids (asteroids with aphelia, 0.73 au$<$ $Q$ $<$ 0.98 au) and \an asteroids \citep[asteroids that have aphelia $Q$ $<$ 0.718 au, e.g.,][]{Bolin2021VR3,Bolin2022IVO}. On average, $\sim$500 sq. deg. is covered during each twilight session to a limiting r-band magnitude of $\sim$20 as described in Section~S1 of \citet[][]{Bolin2023E3sup}. Each field is imaged 4-5 times with 30 s r-band exposures at an airmass of 2-2.5 with a spacing of $\sim$3-5 minutes between each field. A coverage map of the P48/ZTF pointings from 2022 Feb 11-2022 Apr 30 is presented in Fig.~S1 of \citet[][]{Bolin2023E3sup}.

Between 2022 February and 2022 April, $\sim$100,000 sq. deg. of the sky was covered during evening and morning twilight by P48/ZTF as described in Section~S1 of\citet[][]{Bolin2023E3sup}. E3 was discovered on 2022 March 2 in 5 x 30 s r-band P48/ZTF images taken at an airmass of $\sim$2.3 and seeing of $\sim$2\arcsec as measured in the images. A mosaic of the discovery detections of E3 are displayed in Fig.~1.

\begin{figure}\centering
\includegraphics[width=.45\linewidth]{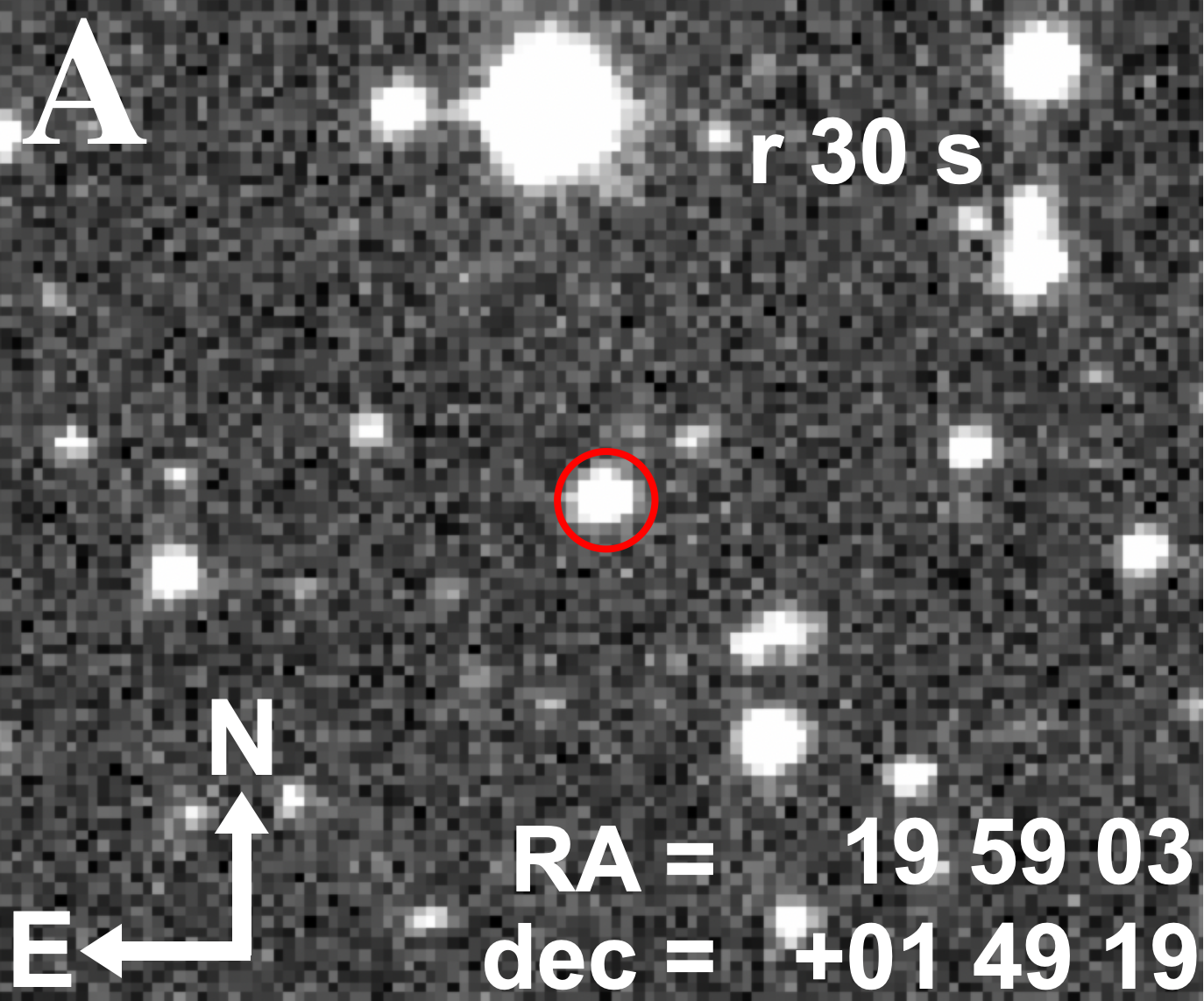} 
\includegraphics[width=.45\linewidth]{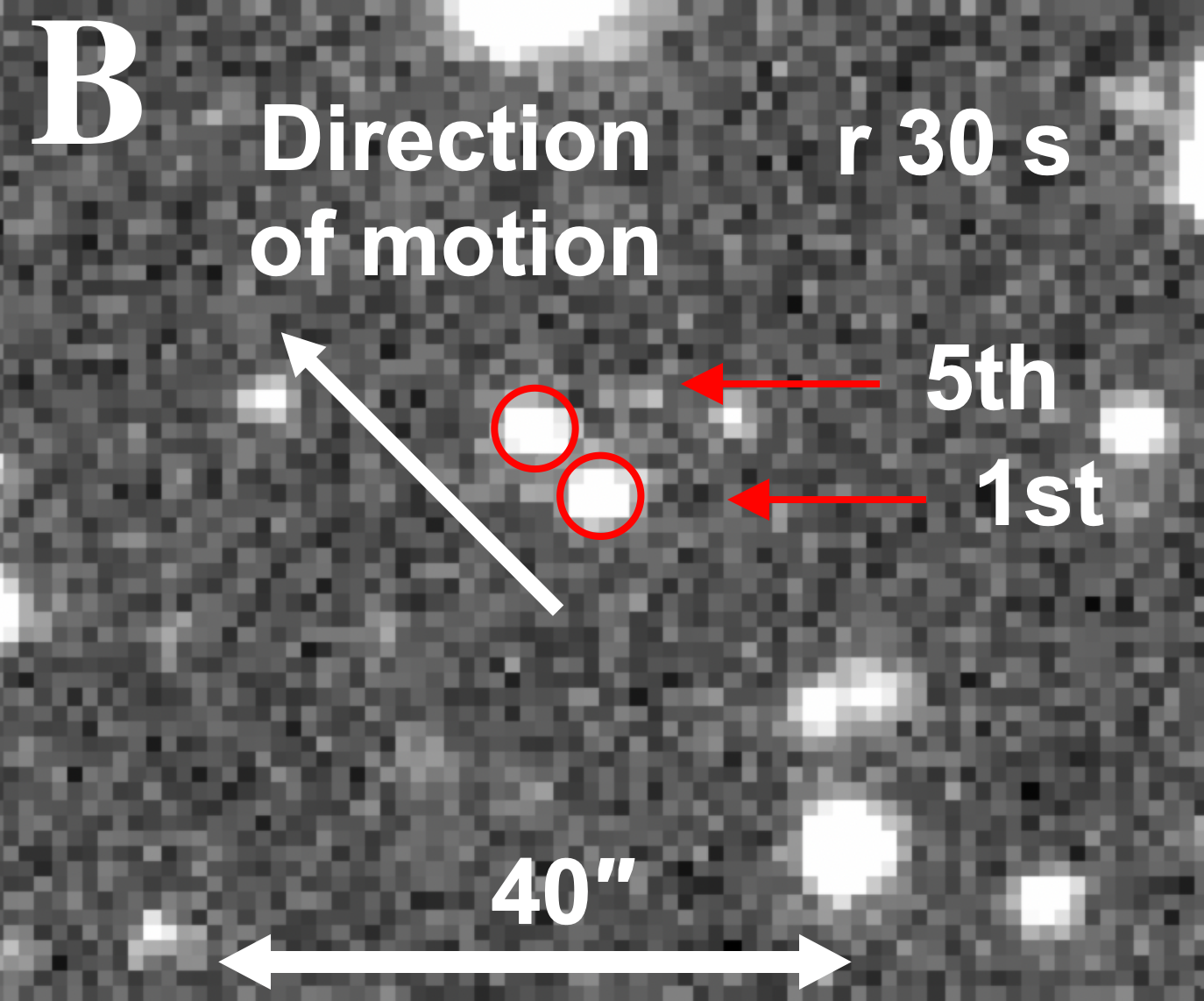}
\includegraphics[width=.45\linewidth]{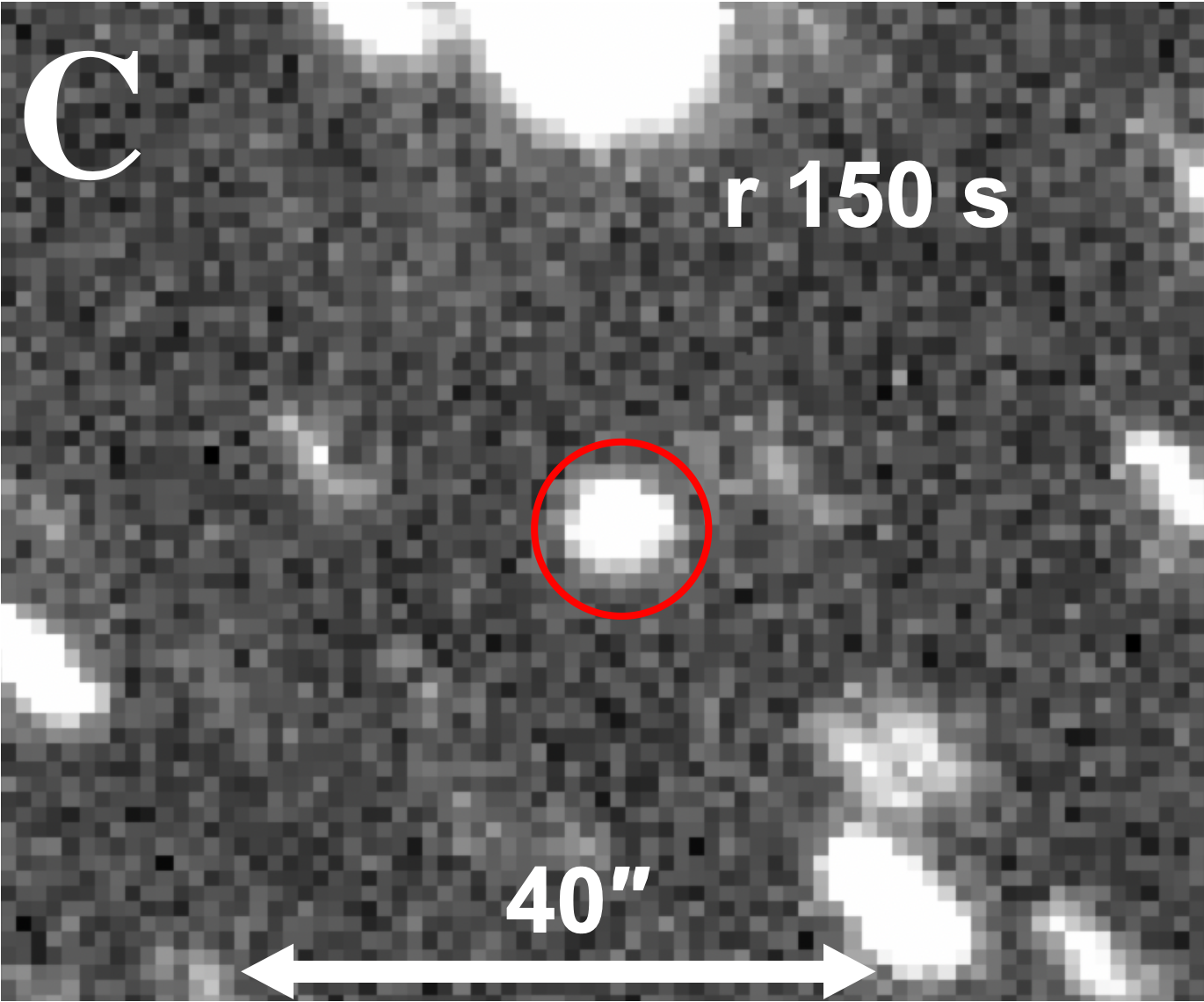} 
\caption{\textbf{P48/ZTF images of E3 taken on 2022 March 2}. Panel (A): The first of the five  30~s r-band P48/ZTF discovery images of E3 taken on 2022 Mar 2. The detection of E3 is encircled in red. Panel (B): Composite image showing the position of E3 in the first and fifth of the 5 discovery images separated by $\sim$15 minutes. The images were aligned on the background stars before being coadded. The first and fifth images are labeled. The comet was moving $\sim$28\arcsec/h in the northeast direction separating the first and fifth detection of E3 by 6\arcsec. Panel (C): a composite stack of the detections of E3 in all five r-band images with an equivalent total exposure time of 150 s. The detection of E3 does not have an immediately apparent extended appearance, however, the measured FWHM of the comet's detection $\sim$3\arcsec compared to the $\sim$2\arcsec FWHM of nearby background stars of similar brightness. The cardinal directions are indicated in Panel (A). Panels (B) and (C) have the same orientation as in Panel (A). The spatial scale indicated in Panels (B) and (C) are is the same as in Panel (A).}
\end{figure}

During the year following its discovery, E3 was routinely monitored for evolution in its activity \citep[e.g.,][]{Jehin2022E3a,Jehin2022E3b,Jehin2023E3a}. In early 2023 February, the comet came within 0.28 au of the Earth, reaching a peak visual magnitude of V$\sim$5, providing an excellent opportunity to observe its coma and volatile contents in high detail. On 2022 Feb 1, optical imaging of E3 was obtained with the P200 using the WaSP wide-field imager. A series of 8 x 45 s images in g-band were taken when the comet was at an airmass of $\sim$1.1, the seeing was $\sim$2.2\arcsec as measured in the images, and the telescope was tracked at the motion of the comet ($\sim$17\arcsec/min). Similar to the procedures described in \citet[][]{Bolin2020CD3} and \citet[][]{Bolin2020HST}, the images were detrended and stacked on the motion of the comet. The composite image stack of E3 taken with the P200 is shown in Fig.~2.

\begin{figure}
\centering
\includegraphics[width=1.0\linewidth]{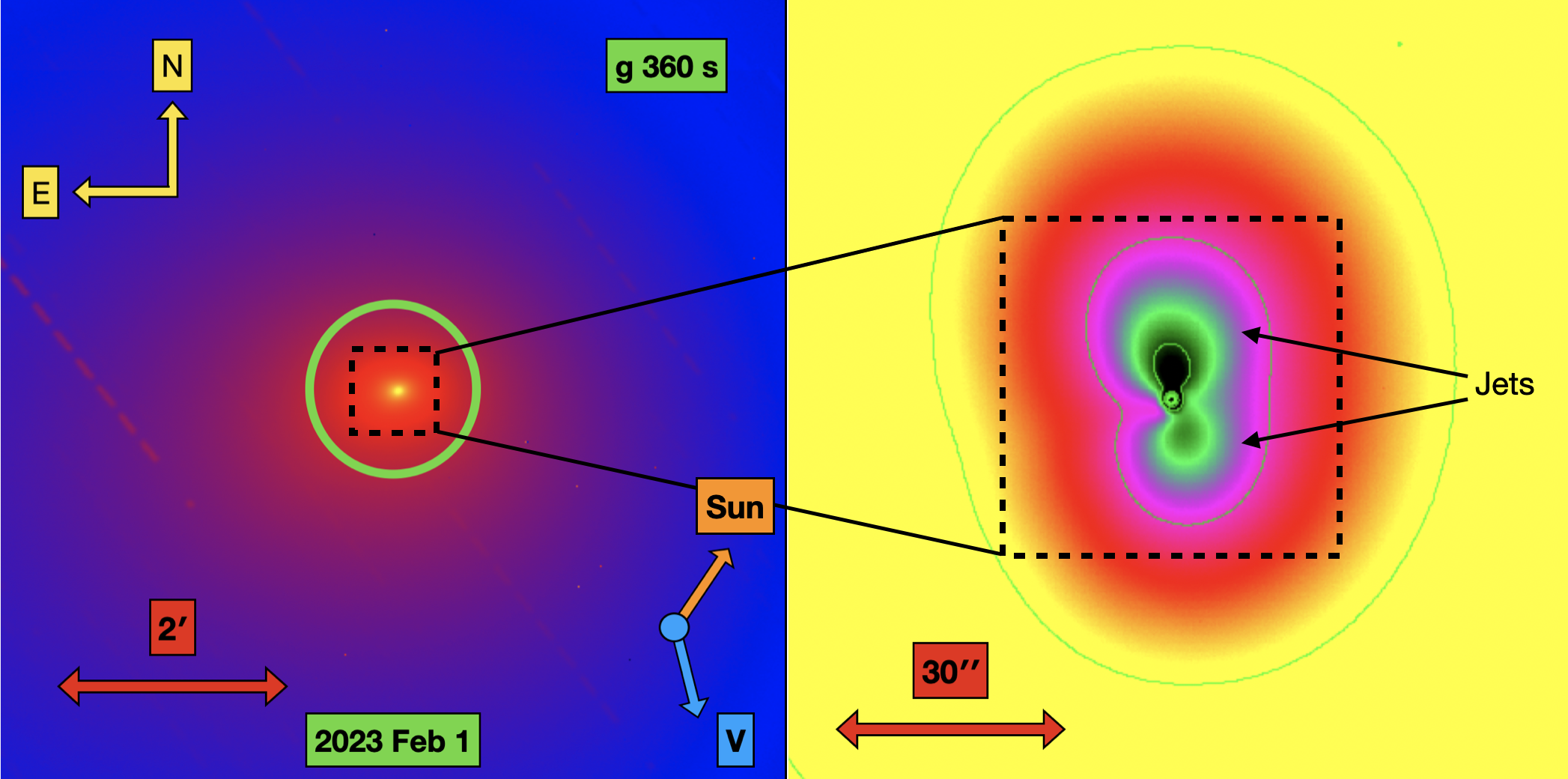}
\caption{\textbf{P200/WaSP g-band imaging of E3 taken on 2023 Feb 1.} Left panel: a composite stack of 8 x 45 s g-band images of E3. The cardinal directions, solar direction, orbital motion direction, and image scale (2\arcmin $\simeq$24,000 km at the 0.28 au distance of the comet) are indicated. A logarithmic colour scale is used where yellow corresponds to the brightest pixels and blue corresponds to the faintest pixels. Right panel: Zoom in on the inner 45\arcmin nucleus region of E3 normalized by the distance from the optocenter to enhance features in the near-nucleus coma. The location of jet features are indicated with black arrows. A logarithmic colour scale is used where black corresponds to the brightest pixels and yellow corresponds to the faintest pixels.}
\end{figure}

 The SpeX instrument on IRTF was used to obtain near-infrared spectra of E3 2023 Feb 26 when the comet was 0.85 au from the Earth and V$\sim$6.5 as reported in observations submitted to the Minor Planet Center (MPC)\footnote{\url{https://minorplanetcenter.net/db_search/show_object?
 =&object_id=C\%2F2022+E3}}. The 0.8\arcsec x 15\arcsec slit was used providing a R$\sim$200 spectrum of the comet between 0.7-2.5~$\mu$m. The comet was observed at an airmass of $\sim$1.7, and a nearby G-type star with colours similar to the Sun's was taken at a similar airmass as E3 to be used for telluric and slope correction \citep[][]{Lewin2020}. The seeing was $\sim$1\arcsec as measured in the trace at 1.7~$\mu$m in the trace of the standard star. 

Similar to the procedures described in \citet[][]{Bolin20202I}, the spectrum was taken as two ABBA sequences with each exposure possessing an equivalent exposure time of $\sim$135 s for a total integration time of 1080 s. The spectrum of E3 was extracted with a 2.4\arcsec wide trace and was divided by the solar analog for telluric and slope correction resulting in the reflectance spectrum shown in Fig.~3.

\begin{figure}
\centering
\includegraphics[width=1.0\linewidth]{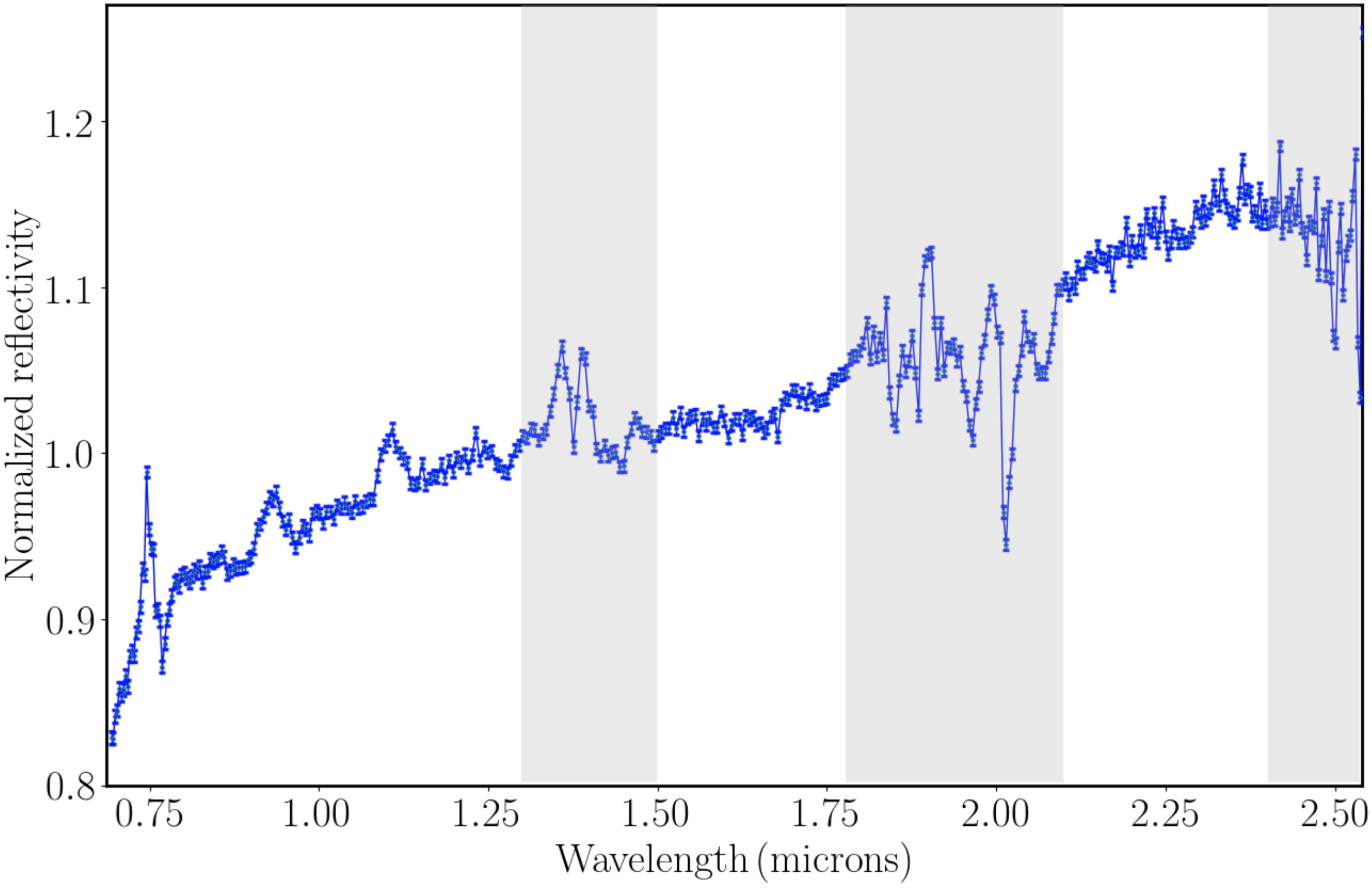}
\caption{\textbf{IRTF/SpeX near-infrared reflectance spectrum of E3 taken on 2023 Feb 26.} The spectrum of E3 has been corrected for telluric features and for the solar spectrum by dividing it by a spectrum of a nearby solar analog star and rebinned by a factor of two. The spectrum has been normalized to unity at 1.25~$\mu$m. The features at 0.75~$\mu$m, 0.90~$\mu$m, and 1.10~$\mu$m are caused by imperfect removal of telluric features. Regions in the spectrum with low atmospheric transparency are highlighted in grey \citep[e.g.,][]{Avdellidou2022}. The error bars correspond to 1-$\sigma$ uncertainties and have been estimated by using the level of scatter in the spectrum \citep[][]{Holler2022}. There is no sign of any water ice absorption feature at $\sim$1.5-1.6~$\mu$m \citep[c.f.,][]{Yang2009,Protopapa2014} down to the $\sim$1$\%$ level.}
\end{figure}

The P60/SEDM  was used to observe E3 on 2023 Mar 10 when the comet was 1.21 au from the Earth and had a reported brightness of V$\sim$7.5. A single 90 s exposure was taken at airmass 1.9 with 2.3\arcsec seeing measured in r-band image portion of the SEDM detector plane and a 7.5\arcsec extraction radius was used providing an R$\sim$100 spectrum of the comet. Similar to the procedures described in \citet[][]{Bolin2023Dink}, a nearby G-type star was taken for telluric and slope correction. The G-type divided spectrum of E3 is presented in Fig.$\sim$3. A detailed description of the P48, P200, IRTF, and P60 observing circumstances are listed in Table~S2 of \citet[][]{Bolin2023E3sup}.

\begin{figure}
\centering
\includegraphics[width=1.0\linewidth]{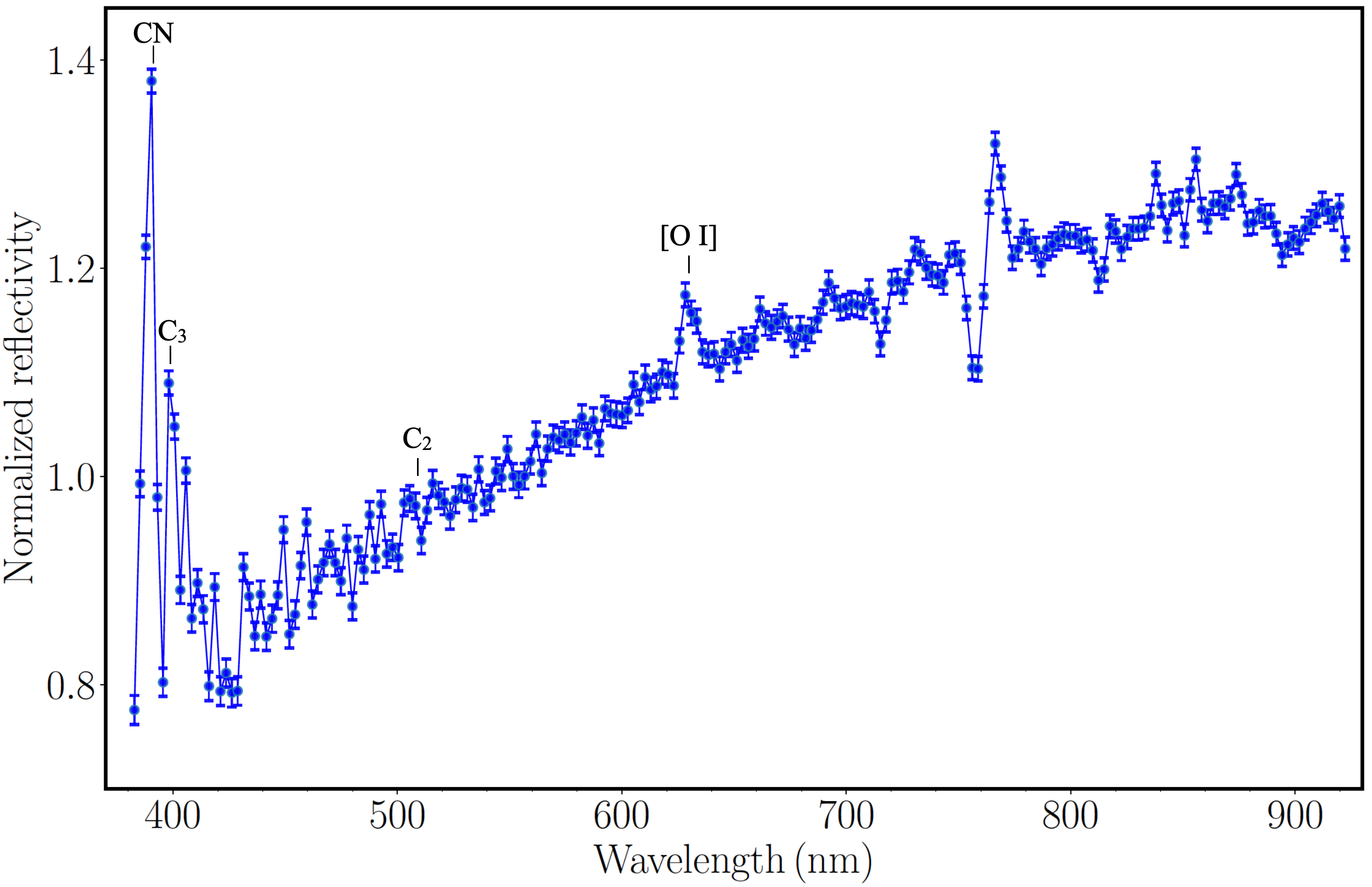}
\includegraphics[width=1.0\linewidth]{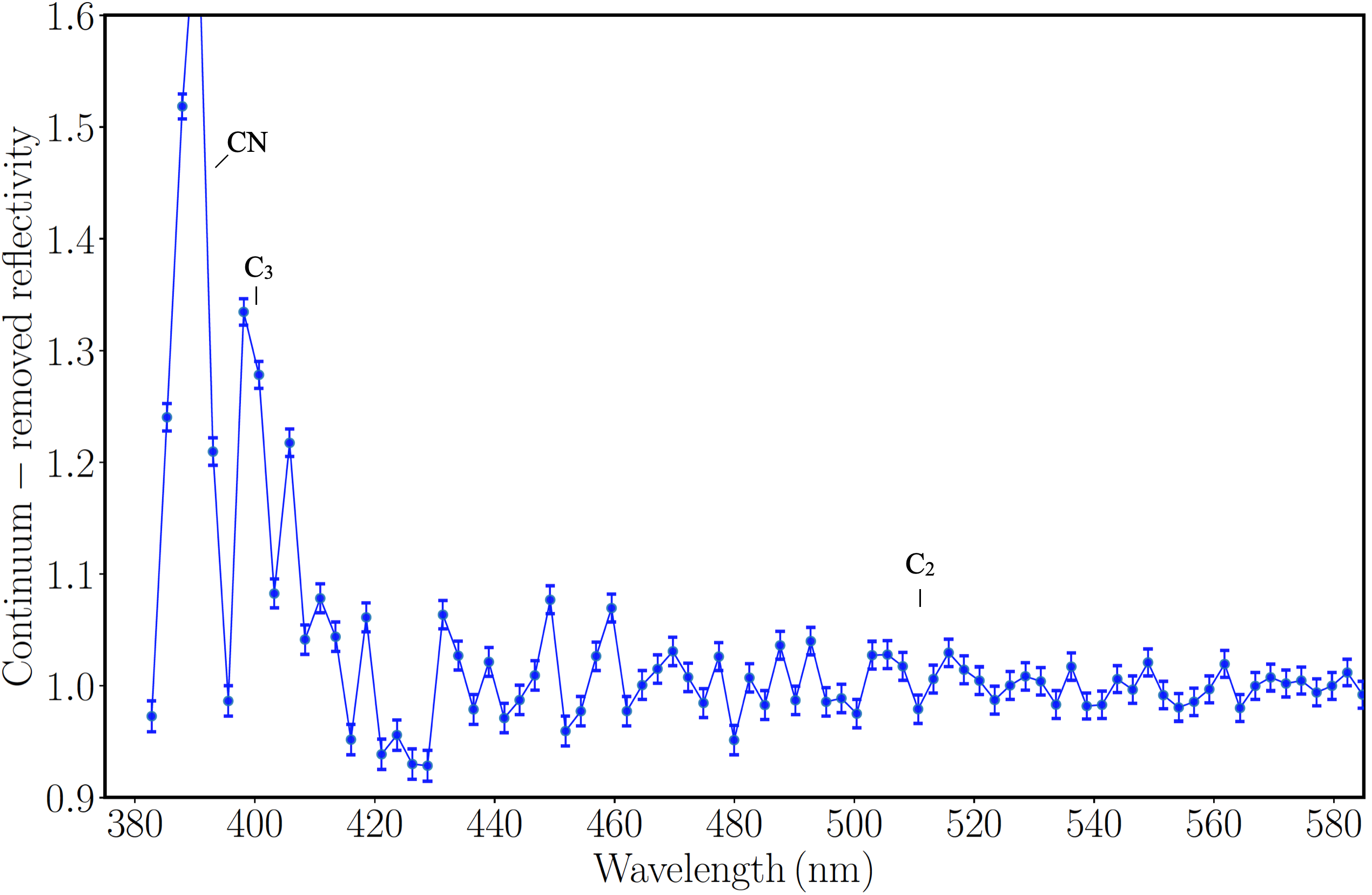}
\caption{\textbf{P60/SEDM visible spectroscopy of E3 on 2023 March 10}. Top panel: visible reflectance spectrum of E3 taken with the P60/SEDM on 2023 Mar 10. The reflectance spectrum was computed by dividing the spectrum of E3 by a local solar analog star to correct for telluric features and the solar spectrum.  The spectrum has been normalized to unity at 550 nm. Bottom panel: a zoomed-in portion of the continuum-removed visible reflectance of E3. The error bars correspond to 1-$\sigma$ uncertainties and have been estimated by using the level of scatter in the spectrum \citep[][]{Holler2022}. The known locations of common cometary volatile species emissions detectable in visible spectra are indicated by small vertical black lines \citep[][]{Farnham2000}. The lowness of the data point at 510 nm compared to adjacent bins may be due to a hot pixel.}
\end{figure}

\section{Results}
\subsection{Initial detection}
C/2022 E3 was detected by P48/ZTF in 5 x 30 s r-band exposures taken in the morning twilight sky $\sim$44$^{\circ}$ from the Sun at  2022 Mar 2 (Fig.~S1 of \citet[][]{Bolin2023E3sup}). Follow-up observations providing additional astrometry extending the orbital arc for E3 were taken by volunteers from the worldwide comet follow-up community \citep[][]{Bolin2022E3}. E3 had a slightly extended PSF with a FWHM of $\sim$3\arcsec in a composite stack of all 5 x 30 s r-band exposures compared to the FWHM of nearby background stars of $\sim$2\arcsec (panel (C) of Fig.~1). The extendedness of E3 was flagged as a possible comet by the Tails neural network trained to identify comets in P48/ZTF images taken on subsequent nights following the discovery \citep[][]{Duev2021}. The activity of the comet was also reported by follow-up observers \citep[][]{Sato2022E3}.

The combined set of observations of E3 including the P48/ZTF discovery images and the follow-up observations refined the orbit revealing that the comet has an eccentricity, $e$, of $\gtrsim$1, a perihelion distance, $q$, of $\sim$1.11 au (Fig.~5) and would reach perihelion on 2023 January 12 and approach within 0.28 au of the Earth in early 2023 February. A complete table of the parameters for the comet's orbital solution including observations taken between 2021 October and 2023 July are listed in Table S1 of \citet[][]{Bolin2023sup}.

\begin{figure}
\centering
\includegraphics[scale = .29]{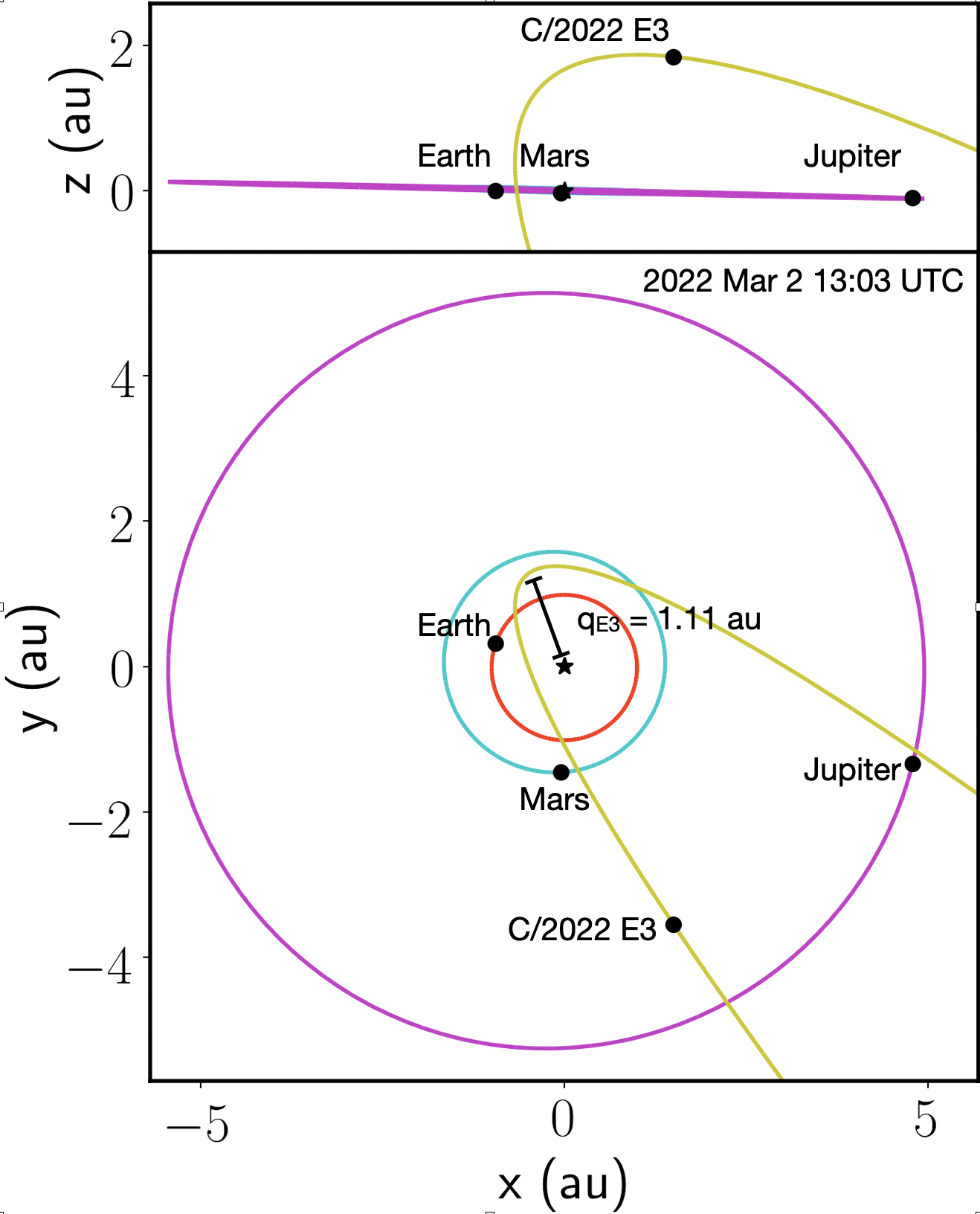}
\caption{\textbf{Orbital configuration of E3 and the inner Solar System on 2022 Mar 2.} Top panel: a side view snapshot of the plane of the solar system with the orbits and locations of Earth (red), Mars (blue), Jupiter (purple), and E3 (green) on 2022 Mar 2 are shown looking from the side. Bottom panel: the same as the top panel but looking from above the orbital plane of the inner solar system. The perihelion distance of E3, q$_{E3}$ is indicated.The heliocentric cartesian coordinates $x$, $y$ and $z$ are indicated with the position of the Sun as the origin.}
\end{figure}

The P200/WasP g-band images of E3 taken on February 1 show the near-nucleus environment of the comet's coma (Fig.~2). The images show the near-nucleus coma as having a width of $\sim$4\arcmin ($\sim$50,000 km at the 0.29 au distance of the comet) and is offset from the nucleus in the anti-solar direction (left panel of Fig.~2). An enhanced version of the g-band image in the right panel of Fig.~2 shows the presence of two jet-like structures in the image indicated with black arrows. One of the jet-like structures points north with a position angle of $\sim$0$^{\circ}$ and one points south with a position angle of $\sim$180$^{\circ}$. 

Both jet-like structures are as long as $\sim$20\arcsec ($\sim$4,200 km at the 0.29 distance of the comet). The jet structures have a broad fan-like shape and may be due to cometary gases jetting from the nucleus as recent observations taken in narrow band comet gas filters have indicated the presence of significant amounts of CN, C$_3$, and C$_2$ in E3 \citep[][]{Jehin2022E3a,Jehin2022E3b,Jehin2023E3a}. The wavelengths between 390 nm and 540 nm of the g-band filter \citep[][]{Fukugita1996} include the emission features from CN, C$_3$, and C$_2$ gases as well as the comet's dust continuum \citep[][]{Farnham2000}. Therefore, the jet features seen in the g-band images may include both cometary gases and dust. However, similar jet-like structures have been detected in narrow-band filters with little dust contamination \citep[][]{Knight2023,Manzini2023}. 

The visible spectrographic observations of E3 were made using the P60/SEDM on 2023 Mar 10 shows a reddish slope with colours at central wavelength locations equivalent to the central wavelength locations at g-band (wavelength $\sim$470~nm), r-band (wavelength $\sim$620~nm), i-band (wavelength $\sim$750~nm), and z-band (wavelength $\sim$900~nm)of g-r = 0.70$\pm$0.01, r-i = 0.20$\pm$0.01, and i-z = 0.06$\pm$0.01 (top panel of Fig.~4). The optical colours of E3 are broadly consistent with the optical broadband colour measurements of active comets made by ground-based surveys \citep[e.g., g-r = 0.5 to 0.7, r-z = 0.2 to 0.5][]{Solontoi2012}. The reddish trend seen in the visible SEDM spectrum continues into the near-infrared in the SpeX spectrum of E3 taken on on 2023 Mar 10. The E3 SpeX spectrum has colours at central wavelength locations equivalent to z-band (wavelength $\sim$900~nm), J-band (wavelength $\sim$1250~nm), H-band (wavelength $\sim$1650~nm), and K-band (wavelength $\sim$2200~nm)  filters of z-J = 0.90$\pm$0.01, J-H = 0.38$\pm$0.01 and H-K = 0.15$\pm$0.01. The visible and near-infrared colours of E3 are significantly redder than the Sun's colours \citep[g$\mathrm{_{\odot}}$-r$\mathrm{_{\odot}}$ = 0.46$\pm$0.01, r$\mathrm{_{\odot}}$-i$\mathrm{_{\odot}}$ = 0.12$\pm$0.01, i$\mathrm{_{\odot}}$-z$\mathrm{_{\odot}}$ = 0.03$\pm$0.01, z$\mathrm{_{\odot}}$-J$\mathrm{_{\odot}}$ = 0.83$\pm$0.01, J$\mathrm{_{\odot}}$-H$\mathrm{_{\odot}}$ = 0.35$\pm$0.01 and H$\mathrm{_{\odot}}$-K$\mathrm{_{\odot}}$ = 0.05$\pm$0.01,][]{Willmer2018}.
 
 The visible SEDM spectrum shows gas emission bands such as CN at $\sim$390 nm, C$_3$ at $\sim$400 nm, and C$_2$ at 510 nm as seen in the top and bottom panels of Fig.~4. The minimum value in the spectrum at 510 nm may be due to a hot pixel in the SEDM IFU. An emission feature possible is detected at $\sim$630 nm possibly corresponding to $[$O I$]$. However, this feature is likely blended with poorly subtracted telluric lines given the limited resolution of our spectrum \citep[c.f.,][]{McKay2020}.

Combining the fluorescence scattering efficiency factors for CN \citep[][]{Schleicher2010}, C$_3$ and C$_2$ \citep[][]{Cochran1992} with an optically thin Haser model \citep[][]{Haser1957} to calculate gas production rates based on the scale lengths derived from \citep[][]{Cochran1992} and an outflow velocity of 0.5 km/s at 1 au \citep[][]{Cochran1993}. We numerically integrated the Haser model over the size of our SEDM aperture of 7.5\arcsec to find production rates for CN of 5.43$\pm0.11\times$10$^{25}$~mol/s, C$_3$ of 2.01$\pm0.04\times$10$^{24}$~mol/s, and C$_2$ of 3.08$\pm0.5\times$10$^{25}$~mol/s. In addition, we calculated a value for A(0$^{\circ}$)f$\rho$,  Af$\rho$corrected to 0$^{\circ}$ phase angle \citep[][]{AHearn1984}, of 1483$\pm$40~cm based on the observed flux within the 7.5\arcsec radius used to extract the SEDM spectrum at $\sim$0.8~$\mu$m. Our gas production rates and A(0$^{\circ}$)f$\rho$ value is similar to the gas production rates and A(0$^{\circ}$)f$\rho$ calculated from independent observations of E3 taken around the same time as our observations \citep[e.g.,][]{Jehin2023E3a}
 
\section{Discussion and conclusion}
E3 seems like an ordinary long-period comet with a red color in the visible and near-infrared with common gas emission features seen in other comets entering the inner solar system. The ratio of C$_2$/CN ($\sim$0.6) is on the lower side compared to most solar system comets with only $\sim$60$\%$ having a higher C$_2$/CN ratio than E3 \citep[][]{AHearn1995}. Comets with C$_2$/CN $<$ 0.6 are considered to be depleted, with the majority of short-period comets, $\sim$40$\%$ having C$_2$/CN $<$ 0.6 \citep[][]{Cochran2012}. Long-period comets, on the other hand, are more likely to be non-depleted, with $\sim$80$\%$ of long-period comets having C$_2$/CN $<$ 0.6 \citep[][]{Cochran2012}. Thus, while having a C$_2$/CN ratio that is relatively lower than most solar system comets, E3 has a C$_2$/CN comparable to many other long-period comets. Additionally, the C$_3$/CN ($\sim$10$^{-1.4}$) ratio  is fairly typical compared to other solar system comets which have  C$_3$/CN $\sim$10$^{-2}$ to 10$^{-1}$ \citep[][]{AHearn1995}. The value for A(0$^{\circ}$)f$\rho$ of 1483$\pm$40~cm is also comparable to other comets at similar solar distances of $\sim$1 au \citep[$\sim$10-10,000 cm,][]{AHearn1995}.

The discovery of E3 is surprising in the sense that its close approach with the Earth in 2023 Jan to 2023 March at solar elongations between 80-110$^{\circ}$ and near naked-eye brightness of V$\sim$5 enabled detailed characterization of its volatile contents and physical properties at viewing geometries accessible to many ground and space-based observatories. The study of long-period comets that brighten dramatically as they pass near the Earth is new territory that will be enhanced by the opportunities for discovering fainter comets at farther heliocentric distances compared to contemporary surveys by near-future survey observatories such as Rubin Observatory's Legacy of Space and Time \citep[][]{Schwamb2023ApJS}. Additionally, the Comet Interceptor spacecraft mission will be designed to rendezvous with comets like E3 pairing well with the expected increasing flux of comet discoveries by future observational surveys \citep[][]{Marschall2022}.

\section*{Acknowledgements}
We wish to recognize and acknowledge the cultural role and reverence that the summit of Maunakea has always had within the indigenous Hawaiian community. The authors wish to recognize and acknowledge the cultural significance that Palomar Mountain has for the Pauma Band of the Luise\~{n}o Indians. Based on observations obtained with the Samuel Oschin Telescope 48-inch and the 60-inch Telescope at the Palomar Observatory as part of the Zwicky Transient Facility project. ZTF is supported by the National Science Foundation under Grants No. AST-1440341 and AST-2034437 and a collaboration including current partners. B.T.B. is supported by an appointment to the NASA Postdoctoral Program at the NASA Goddard Space Flight Center, administered by Oak Ridge Associated Universities under contract with NASA.
%%%%%%%%%%%%%%%%%%%%%%%%%%%%%%%%%%%%%%%%%%%%%%%%%%
\section*{Data Availability}
The data underlying this article will be shared on reasonable request to the corresponding author. The ZTF Survey data from 2022 February to 2022 April are available in ZTF Public Data Release 12.

\section*{Supplemental Material}
The supplemental material for this manuscript is available online.

%%%%%%%%%%%%%%%%%%%% REFERENCES %%%%%%%%%%%%%%%%%%

% The best way to enter references is to use BibTeX:

\bibliographystyle{mnras}
\bibliography{neobib} % if your bibtex file is called example.bib

% Alternatively you could enter them by hand, like this:
% This method is tedious and prone to error if you have lots of references
%\begin{thebibliography}{99}
%\bibitem[\protect\citeauthoryear{Author}{2012}]{Author2012}
%Author A.~N., 2013, Journal of Improbable Astronomy, 1, 1
%\bibitem[\protect\citeauthoryear{Others}{2013}]{Others2013}
%Others S., 2012, Journal of Interesting Stuff, 17, 198
%\end{thebibliography}

%%%%%%%%%%%%%%%%%%%%%%%%%%%%%%%%%%%%%%%%%%%%%%%%%%

%%%%%%%%%%%%%%%%% APPENDICES %%%%%%%%%%%%%%%%%%%%%

\renewcommand{\thefigure}{S\arabic{figure}}
\setcounter{figure}{0}
\renewcommand{\thetable}{S\arabic{table}}
\renewcommand{\theequation}{S\arabic{equation}}
\renewcommand{\thesection}{S\arabic{section}}
\setcounter{section}{0}
\cleardoublepage
\setcounter{page}{1}
\renewcommand\thepage{S\arabic{page}}
\section*{Supplemental Material}
\appendix
\renewcommand{\thefigure}{S\arabic{figure}}
\setcounter{figure}{0}
\renewcommand{\thetable}{S\arabic{table}}
\renewcommand{\theequation}{S\arabic{equation}}
\renewcommand{\thesection}{S}
\setcounter{section}{0}
\subsection{Observational details}
\textit{Palomar 48-inch telescope (P48)/Zwicky Transient Facility (ZTF):} The ZTF instrument mounted on the P48 consists of 16 separate 6144~pixel $\times$ 6160~pixel arrays on a single CCD camera. The camera has a plate scale of arcseconds pixel$^{-1}$ with a 47 sq. deg. field of view \citep[][]{Dekany2020}. The data processing pipeline produces images differenced from reference frames and removes or masks most detector artefacts. Transients are extracted from the images and several algorithms are used to identify moving objects and comets \citep[][]{Masci2019, Duev2021}. 

Comet E3 was discovered in images taken on 2022 Mar 2 at 13:03 UTC. The comet was imaged in five separate sidereally-tracked 30~s r-band images (Fig.~1 of the Main Text). The observations were made during morning astronomical twilight while the telescope was pointing at 25.1 elevation and the center of the telescope's field of view was pointing through an airmass of 2.3 and with seeing of $\sim$2\arcsec. The brightness of E3 during its discovery observations was r$\sim$17.3 magnitude and was moving approximately $\sim$28~\arcsec/h. The comet was at this time 4.94 au from the Earth and 4.28 au distant from the Sun.

At the time E3 was discovered, observations during morning twilight were taking place every morning with two sets of fields bimodally offset from the Sun. The first set of fields focused on sky $\sim$45 from the sun while the second set of fields focused on sky $\sim$55 from the Sun. Each of the two sets of fields consisted of 5 x 30 s r-band images separated by a few minutes apart. A map in Sun-centric ecliptic coordinates of the P48/ZTF pointings during 2022 Feb 11 and 2022 Apr 30 during which E3 was discovered is presented in Fig.~S1 of \citet[][]{Bolin2023E3sup}.

\begin{figure}
\hspace{0 mm}
\centering
\includegraphics[scale = .25]{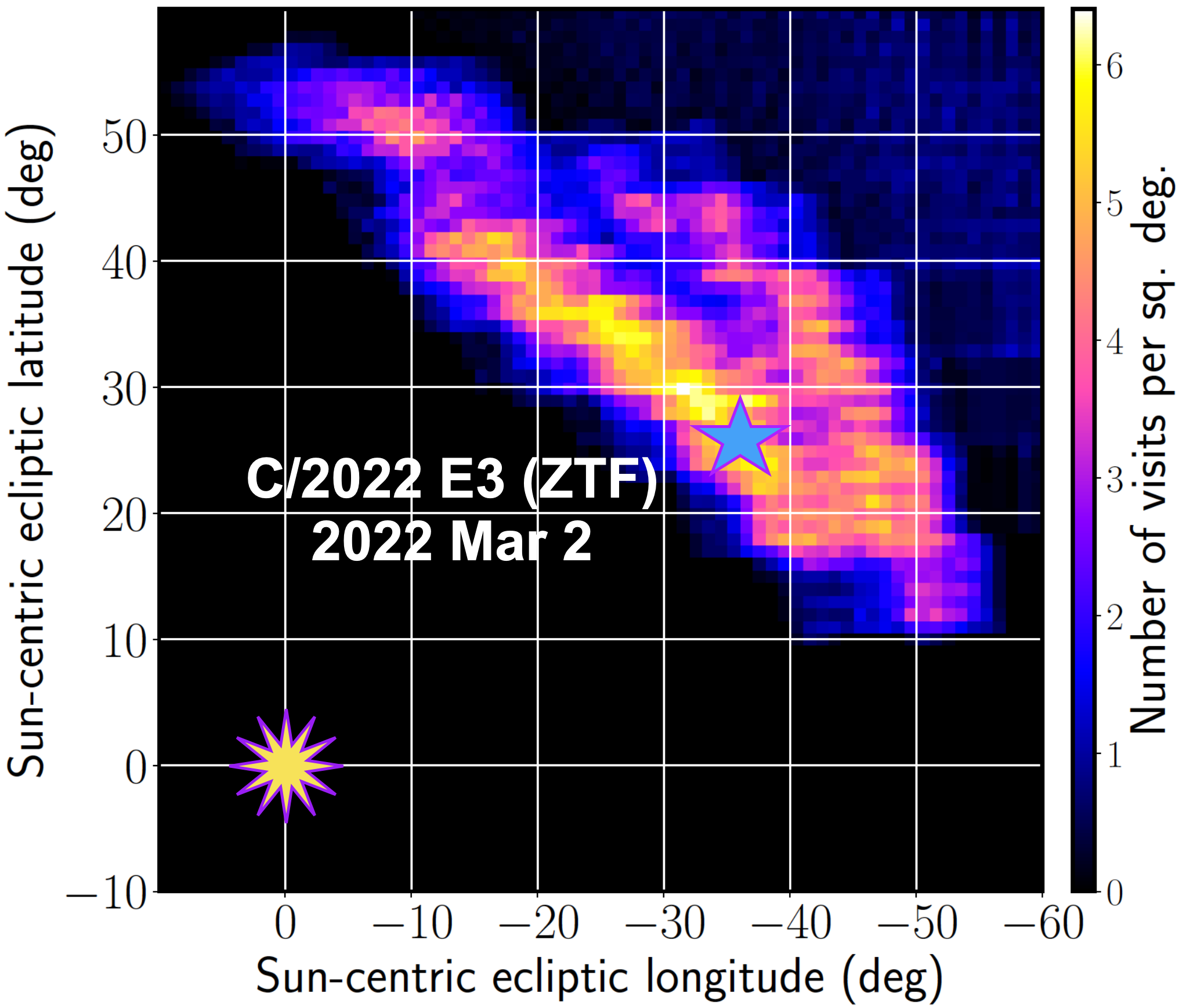}
\caption{\textbf{P48/ZTF coverage of the sky in Sun-centric ecliptic coordinates.} The ZTF sky coverage shown here for dates 2022 Feb 11 and 2022 Apr 30 is in the morning twilight sky. The blue star indicates the observed location of E3 on 2022 Mar 2 and the yellow star indicates the Sun. The colour scale is the number of ZTF visits per sq. deg.}
\end{figure}

The initial P48/ZTF observations of E3 were submitted to the Minor Planet Center (MPC) where it appeared on the Possible Comet Confirmation Page\footnote{\url{https://minorplanetcenter.net/iau/NEO/pccp_tabular.html}}. The discovery of E3 was announced on 2022 Mar 21 after numerous observations confirming the orbit and activity were made by volunteer observers \citep[][]{Bolin2022E3,Sato2022E3}. Additional confirmation of the comet's activity was made in P48/ZTF images following the discovery by the Tails pipeline \citep[][]{Duev2021}. Additional follow-up observations have been reported up to 2023 July and pre-discovery observations have been reported as far back as 2021 Oct\footnote{\url{https://minorplanetcenter.net/db_search/show_object?utf8=&object_id=C\%2F2022+E3}}. The list of orbital parameters for E3 is presented in Table~S1 and a diagram of the orbit is shown in Fig.~5 of the Main Text. 
\\
\begin{table}
\centering
\caption{\textbf{Orbital elements of C/2022 E3 (ZTF) based on observations reported to the MPC taken between 2021 Oct 25 to 2023 July 18.} The orbital elements are shown for the Julian date (JD) 2,459,865.5. The 1-$\sigma$ uncertainties are given in parentheses. The value and 1-$\sigma$ uncertainties for $H$ were taken from the JPL Small-Body Database on 2023 August 24.}
\label{t:hstobs}
\begin{tabular}{ll}
\hline
Element&
\\ \hline
Epoch (JD) & 2,459,865.5\\
Time of perihelion, $T_p$ (JD) & 2,459,957.2842479$\pm$(4.48x10$^{-5}$)\\
Perihelion, $q$ (au) & 1.112238535$\pm$(4.22x10$^{-7}$)\\
Eccentricity, $e$ & 1.000303087$\pm$(6.85x10$^{-7}$)\\
Inclination, $i$ ($^{\circ}$) & 109.1691691$\pm$(1.25x10$^{-5}$)\\
Ascending node, $\Omega$ ($^{\circ}$) & 302.5547345$\pm$(1.14x10$^{-5}$)\\
Argument of perihelion, $\omega$ ($^{\circ}$) & 145.8148729$\pm$(2.58x10$^{-5}$)\\
Mean Anomaly, $M$ ($^{\circ}$) & -0.00040694$\pm$(1.38x10$^{-6}$)\\
Comet Total Magnitude, M$_1$ & 10.8$\pm$(0.8)\\
\hline
\end{tabular}
\end{table}
\\
\noindent\textit{Palomar 200-inch telescope (P200)/Wafer-Scale Imager for Prime (WaSP)} The WaSP instrument mounted at prime focus on the P200 was used to observe E3 on 2023 Feb 1 under program 2022B-P99 (PI: J. Milburn). The WaSP detector array includes 4 x 6144 $\times$ 6160 Teledyne e2v array with a pixel scale of 0.19\arcsec pixel$^{-1}$ \citep[][]{Nikzad2017}. E3 was observed with SDSS g ($\mathrm{\lambda_{eff}}$ = 467.2 nm, FWHM = 126.3 nm) filter 
\citep[][]{Fukugita1996}. A total of 8 x 45 s g-band exposures were taken while the telescope was tracked at the on-sky rate of motion of E3 of 16.6\arcsec/min. The observations were taken at an airmass of $\sim$1.1 in 2.2\arcsec seeing. 
\\
\textit{Infrared Telescope Facility (IRTF)/SpeX:} 
The medium-resolution near-infrared spectrograph SpeX mounted on the IRTF was used to observe E3 on 2023 Feb 26 under program 2023A018 (PI: B. Bolin). SpeX consists of a Teledyne 2048x2048 Hawaii-2RG with a pixel scale of 0.1\arcsec \citep[][]{Rayner2003}. A prism disperser was used with a 0.8\arcsec x 15\arcsec slit providing a 0.7-2.5~$\mu$m spectrum with resolving power of $\sim$200. A total of 8 x 135 s exposures were taken with an ABBA dither pattern with an 8\arcsec spacing between dithers. The observations were taken at an airmass of 1.7 in $\sim$1\arcsec seeing as measured in the images. The wavelength solution was obtained using arcs within the internal calibration unit of SpeX. A spectrum of the solar analog star HD 37685 was taken for telluric and slope correction. Detrending of the data and extraction of the spectra were completed with spextool \citep[][]{Cushing2004}.
\\
\textit{Palomar 60-inch telescope (P60)/Spectral Energy Distribution Machine (SEDM):} The P60/SEDM integral field unit (IFU) spectrograph was used to observe E3 on 2023 Mar 10 under program 2022B-Asteroids (PI B. Bolin). The SEDM observations of E3 occurred when the comet was at an airmass of 1.9 in 2.3\arcsec seeing. SEDM consist of a 28\arcsec x 28\arcsec spaxel array with a spaxel scale of 0.75\arcsec providing a 0.38-0.92~$\mu$m spectrum with a resolving power of $\sim$100 \citep[][]{Blagorodnova2018}. The spectrum of a nearby G-type star, HD 28192, was taken for telluric and slope correction. An extraction radius of 7.5\arcsec was used to extract the spectrum of E3 and the G-type star in the IFU data. The SEDM data for both E3 and HD 28192 were reduced using pysedm \citep[][]{Rigault2019}. SmallBodyPython was used to compute the gas production rates and Af$\rho$ from our visible spectrum \citep[][]{Mommert2019sbpy}.

\begin{table*}
\caption{\textbf{Observational details.}}
\centering
\begin{tabular}{llllllllll}
\hline
Facility & Instrument & UT Date    & $\Delta^1$ & r$_H^2$ & $\alpha^3$     & Seeing$^4$                   & Airmass & Sky motion                 & V$^5$     \\
         &            &            & (au)     & (au)  & ($^{\circ}$) & (\arcsec) &         & (\arcsec/h) & (mag) \\ \hline
P48      & ZTF        & 2022-03-02 & 4.937    & 4.276 & 9.2          & 2.0                      & 2.3     & 28.09                      & 17.5  \\
P200     & WaSP       & 2023-02-01 & 0.285    & 1.155 & 47.5         & 2.2                      & 1.1     & 995.86                     & 5.0   \\
IRTF     & SpeX       & 2023-02-26 & 0.847    & 1.320 & 48.6         & 1.0                      & 1.7     & 108.35                     & 6.5   \\
P60      & SEDM       & 2023-03-10 & 1.207    & 1.428 & 43.3         & 2.3                      & 1.9     & 53.70                      & 7.5  \\ \hline
\end{tabular}
\begin{tablenotes}
\item \textbf{Notes.} (1) Geocentric distance, (2) heliocentric distance, (3) phase angle, (4) measured in science images, (5) V-band magnitude as measured in the P48/ZTF images taken of E3 on 2022 Mar 2 with a 2\arcsec radius aperture. The other entries in this column are taken from observations reported on these dates to the MPC.
\end{tablenotes}
\end{table*}

%%%%%%%%%%%%%%%%%%%%%%%%%%%%%%%%%%%%%%%%%%%%%%%%%%

% Don't change these lines
\bsp	% typesetting comment
\label{lastpage}
\end{document}